# Ce doping in $T$-$La_2CuO_4$ films: Broken electron-hole symmetry for high-$T_c$ superconductivity

A. Tsukada [a,*], H. Yamamoto [a], M. Naito [b]

[a] NTT Basic Research Laboratories, NTT Corporation, 3-1 Morinosato-Wakamiya, Atsugi, Kanagawa 243-0198, Japan

[b] Department of Applied Physics, Tokyo University of Agriculture and Technology, 2-24-16 Naka-cho, Koganei, Tokyo 184-8588, Japan

## Abstract

We attempted Ce doping in $La_2CuO_4$ with the $K_2NiF_4$ ($T$) structure by molecular beam epitaxy. At low growth temperature and with an appropriate substrate choice, we found that Ce can be incorporated into the $K_2NiF_4$ lattice up to $x \sim 0.06$, which had not yet been realized in bulk synthesis. The doping of Ce makes $T$-$La_{2-x}Ce_xCuO_4$ more insulating, which is in sharp contrast to Ce doping in $La_2CuO_4$ with the $Nd_2CuO_4$ structure, which makes the compounds superconducting. The observed smooth increase in resistivity from hole-doped side ($T$-$La_{2-x}Sr_xCuO_4$) to electron-doped side ($T$-$La_{2-x}Ce_xCuO_4$) indicates that electron-hole symmetry is broken in the $T$-phase materials.

## 1, Introduction

The phase diagram of hole-doped high-$T_c$ cuprates shows a well-known 'dome' shape with maximal superconductivity at a doping level of about 0.15, and the currently proposed phase diagram of electron-doped high-$T_c$ cuprates is roughly similar to the hole-doped one [1]. Hence, it has been claimed that "electron-hole" symmetry holds for high-$T_c$ superconductivity. This claim has been one ground supporting the doped Mott insulator scenario [2], in which the parent material is a Mott insulator and high-$T_c$ superconductivity develops with either hole or electron doping into the Mott insulator. However, the argument for the above "electron-hole" symmetry is based on the comparison of hole and electron doping into different structures, namely, hole doping in the $K_2NiF_4$ (so-called *T*) structure [3] and electron doping in the $Nd_2CuO_4$ (*T'*) structure [4]. These two structures are similar in appearance but different in nature. Namely, the difference in the *RE*-O arrangement, rock-salt-like in *T* and fluorite-like in *T'*, produces a significant change in the Cu-O coordination: octahedral $CuO_6$ in *T* and square-planar $CuO_4$ in *T'*. In principle, it is desirable to compare hole and electron doping in the same crystal structure. Such a comparison has not yet been performed since it is empirically known in bulk synthesis that hole doping is possible only in octahedral or pyramidal ($CuO_5$) cuprates, whereas electron doping is possible only in square-planar cuprates. In this article, we report that Ce can be doped up to $x \sim 0.06$ in *T*-$La_2CuO_4$ by a low-temperature thin-film route. The linear decrease of the *c*-axis lattice constant ($c_0$) and in-situ X-ray photoelectron spectroscopy (XPS) indicate that the valence state of Ce in *T*-$La_2CuO_4$ films is close to +4. Hence, effective electron doping seems to have been achieved in octahedral cuprates for the first time. This novel success led us to the unexpected finding that electron doping makes *T*-$La_2CuO_4$ more insulating than in the undoped state, indicating that "electron-hole" symmetry is apparently broken in this compound.

## 2. Experimental

$La_{2-x}Ce_xCuO_4$ thin films were grown in a customer-designed molecular beam epitaxy (MBE) chamber (base pressure ~ $10^{-9}$ Torr) from metal sources using multiple electron-gun evaporators with atomic oxygen (1sccm) activated by RF power of 300 W. The details of our MBE growth are described elsewhere [5]. There are two problems specific to MBE growth of *T*-$La_{2-x}Ce_xCuO_4$ films. The first concerns the inclusion of *T'*-$La_{2-x}Ce_xCuO_4$ as an impurity phase. Because of the low synthesis temperature employed in MBE growth, the *T'* phase is easy to form and even predominates in preparation of $La_{2-x}Ce_xCuO_4$ films for $x$ even as small as 0.05 [6, 7]. *T'*-$La_{2-x}Ce_xCuO_4$ films are more conductive than *T*-$La_{2-x}Ce_xCuO_4$ films by three to six orders of magnitude. Therefore, even a tiny inclusion of the *T'* phase ruins the generic transport properties of the *T* phase. In order to prepare phase-pure *T*-$La_{2-x}Ce_xCuO_4$ films, the substrate choice is crucial. Our previous results on the phase control of undoped $La_2CuO_4$ indicated that $K_2NiF_4$-type substrates have a stronger tendency to stabilize the *T* phase than perovskite substrates [8]. Hence, we used $K_2NiF_4$-type substrates in this work, which are listed in Table I. The second problem in *T*-$La_{2-x}Ce_xCuO_4$ growth is that Ce tends to segregate out from the *T* lattice at growth temperatures higher than 700˚C [9]. The films grown at $T_s$ > 700˚C showed no change in the *c*-axis lattice constant ($c_0$) with the Ce content, indicating that Ce is not incorporated into the lattice. This problem forced us to lower the growth temperature to well below 700˚C, which, in turn, favors the *T'*-phase formation. As a compromise between the first and second problems, we employed $T_s$ = 630°C. The typical film thickness was as thin as 450 Å in order to make full use of the epitaxy effect. After the growth, the films were cooled to ambient temperature in a vacuum of $P_{O2}$ at less than $10^{-9}$ Torr to avoid the introduction of excess oxygen into the films.

The films were characterized by X-ray diffraction and resistivity measurements. All the Ce doped *T*-phase films had very high resistance (typically > 100 MΩ) even at 300 K, so electrodes with low contact resistance were required for reliable measurements. We formed electrodes by Ag evaporation. The sampling time for resistivity measurements had also to be set to several seconds. The valence state of Ce was determined by XPS measurements, for which the films were transferred to a surface analysis chamber in vacuum via a gate valve.

## 3. Results and discussions

Figure 1 compares the X-ray diffraction patterns of $La_{2-x}Ce_xCuO_4$ film grown on $NdCaAlO_4$ (NCAO), $LaSrAlO_4$ (LSAO), and $LaSrGaO_4$ (LSGO) substrates with Ce concentration $x$ = 0.06. Since the $c_0$ values of *T*- and *T'*-structure are distinct, the phase identification is rather straightforward [8]. The film grown on NCAO substrate is single-phase *T'*, the film on LSAO is a mixture of *T*- and *T'*-phase, and the film on LSGO is single-phase *T*. The same experiments were performed for different $x$, and the results are summarized in Fig. 2, which shows a phase diagram of the grown phases as a function of $x$ for each substrate. Here, the $a$-axis lattice constant ($a_s$) of the substrates is taken as the vertical axis. The films with $x$ = 0 are single-phase *T* on all substrates. With increasing $x$, one can observe a transition from *T* to *T'*. The critical doping ($x_{T\text{-}T'}$) for this transition is substrate-dependent. On NCAO, which has the smallest $a_s$ (3.688 Å) of the substrates investigated in this work, $x_{T\text{-}T'}$ is 0.015 ~ 0.03, whereas on LSGO, which has the largest $a_s$ (3.843 Å), $x_{T\text{-}T'}$ is 0.06 ~ 0.09. There is a clear trend that larger $a_s$ substrates stabilize *T*-structure until higher *x* is reached. This trend can be understood as follows. Assuming that Ce doping provides electrons effectively for the copper oxide plane, electron doping should stretch Cu-O bonds by filling electrons in the antibonding $dp\sigma$ orbitals and thereby expands the in-plane lattice constant ($a_0$) [4, 10]. Hence, $La_{2-x}Ce_xCuO_4$ films with higher *x* match in lattice with larger $a_s$ substrates.

Figure 3 plots the $c_0$ value of resultant $La_{2-x}Ce_xCuO_4$ films grown on $LaSrGaO_4$ substrates against *x*. At $x = 0.09$, where the film is a two phase mixture of *T* and *T'*, both of the $c_0$ values for *T*- and *T'*-phase are plotted. In the *T*-phase, $c_0$ decreases linearly with increasing *x*. This seems to indicate that the substitution of La by Ce in the *T*-structure is successful up to $x = 0.09$. Although the $c_0$ value shows a systematic change with the Ce content, the in-plane lattice constant ($a_0$) remains almost unchanged with doping and is equal to the substrate $a_s$. This is due to the epitaxial-strain effect. As we have demonstrated before, the in-plane lattice constant of *T*-$(La,Sr)_2CuO_4$ films on $LaSrAlO_4$ substrates also show only a small change with Sr doping and is almost equal to the $a_s$ of $LaSrAlO_4$ [11, 12].

Figure 4 shows the temperature dependence of resistivity ($\rho$) for the films grown on $LaSrGaO_4$ substrates with different *x*. As seen in the figure, the *T*-phase films become more insulating with increasing *x*. The metallic behavior with a superconducting transition at about 30 K observed in the film with $x = 0.09$ is due to the impurity phase *T'*-$La_{2-x}Ce_xCuO_4$ [6, 7]. Figure 5 plots the resistivity at 300 K [$\rho$(300 K)] as a function of the Ce content. It also includes our data for *T*-$La_{2-x}Sr_xCuO_4$ (hole doping into the *T*-structure) [12, 13] and *T'*-$La_{2-x}Ce_xCuO_4$ (electron doping into the *T'*-structure) [7, 14]. The $\rho$(300 K) of *T*-$La_{2-x}Ce_xCuO_4$ films increases with Ce doping, whereas that of *T*-$La_{2-x}Sr_xCuO_4$ films shows a rapid decrease with Sr doping. Taking a look at the global variation of the resistivity from the hole (Sr) doped side to the electron (Ce) doped side in *T*-$La_2CuO_4$, the resistivity at 300 K starts to rise rapidly with a break in the slope of $\rho$-vs-$x$ at hole doping of about 0.05 and steadily increases up to electron doping of about 0.06 with no break at $x \sim 0$. In contrast, Ce doping in *T'*-$La_2CuO_4$ lowers $\rho$(300 K) gradually.

The key question is the valence state of Ce dopants in *T*-$La_2CuO_4$. It is known that Ce can take a trivalent as well as tetravalent state. If Ce *were* trivalent, it *would* not be surprising that Ce doping leaves *T*-$La_2CuO_4$ insulating. The evaluation of the Ce valence is difficult by any wet chemical analysis for thin films, so it was performed by *in-situ* XPS. The result indicates that the valence of Ce dopants in *T*-$La_2CuO_4$ is nearly +4 [15]. This is also supported by the doping dependence of $c_0$. The slope of $c_0$ with $x$ for *T*-$La_{2-x}Ce_xCuO_4$ films is similar to that for superconducting *T'*-$La_{2-x}Ce_xCuO_4$ films [14], indicating that the ionic size of Ce dopants is close to $Ce^{4+}$ (0.97 Å) rather than $Ce^{3+}$ (1.196 Å) in either compound [16]. Hence, these results suggest that effective electron doping in *T*-$La_2CuO_4$ is achieved by tetravalent Ce doping. In addition, the substantial substrate dependence of the *T*-phase stability shown in Fig. 2 is also strong evidence that electron doping is actually achieved. If it is, the resistivity data in Fig. 5 indicates that electron-hole symmetry is apparently broken in this compound.

Next, one has to inquire as to the nature of the insulating state in electron-doped *T*-$La_2CuO_4$. If *T*-$La_2CuO_4$ *were* a Mott insulator, electron doping *would* lead to a metallic state. But our experimental results are just the opposite. Then, one has to think of other possible insulating mechanisms. One simple possibility is to regard the insulating antiferromagnetic ground state in $T$-$La_2CuO_4$ as a Fermi-surface driven spin-density-wave (SDW) state. Then, the nesting of Fermi surface could occur in a larger area at finite doping than at zero doping, depending on the shape of the Fermi surface. Another possibility is to regard $T$-$La_{2-x}Ce_xCuO_4$ as a Kondo insulator. In this scenario, strong hybridization between O2$p$ holes and Cu3$d$ electrons leads to a large Kondo coupling ($J_K$). Hence, each O2$p$ hole is strongly bound to Cu3$d$ spins, forming a localized Kondo singlet (so-called Zhang-Rice singlet) with a small energy gap. One calls the resultant small-gap semiconductors Kondo insulators. The latter possibility seems to us to be more likely [17, 18].

## 4. Summary

We succeeded in the growth of $T$-$La_{2-x}Ce_xCuO_4$ with $x \leq 0.06$ by molecular beam epitaxy. All $T$-$La_{2-x}Ce_xCuO_4$ films are insulating and become more insulating with electron doping. The valence of Ce in these materials is nearly +4, which means electron doping is actually achieved. Our finding indicates that electron-hole symmetry is broken in $T$-$La_2CuO_4$ and also that the insulating nature in $T$-$La_{2-x}Ce_xCuO_4$ is of the Kondo insulator instead of that of the Mott insulator.

## Acknowledgments

The authors thank Dr. T. Yamada, Dr. A. Matsuda, Dr. H. Sato, Dr. S. Karimoto, Dr. K. Ueda, and Dr. J. Kurian for helpful discussions, and Dr. T. Makimoto, Dr. K. Torimitsu, Dr. M. Morita, Dr. H. Takayanagi, and Dr. S. Ishihara for their support and encouragement throughout the course of this study.

## Figure captions

Fig. 1. Typical X-ray diffraction pattern of $La_{2-x}Ce_xCuO_4$. Films were grown on (a) $NdCaAlO_4$, (b) $LaSrAlO_4$, and (c) $LaSrGaO_4$ substrates with $x = 0.06$. The substrate peaks are removed.

Fig. 2. Phase diagram of the selective stabilization of *T* versus *T'* in the $a_s$-$x$ plane. Closed circles and squares represent single-phase *T* and *T'*, respectively. Open triangles represent a mixture of *T* and *T'*.

Fig. 3. The $c$-axis lattice constants ($c_0$) of $La_{2-x}Ce_xCuO_4$ films grown on $LaSrGaO_4$ substrates as a function of Ce concentration $x$. Closed circles and square represent $c_0$ of *T*- and *T'*-phase, respectively. The dotted line is guide for the eye.

Fig. 4. Temperature dependence of resistivity for $La_{2-x}Ce_xCuO_4$ films grown on $LaSrGaO_4$ substrates with different $x$. The solid lines are for films of single-phase *T* (●: $x = 0$; ■: 0.015; ♦: 0.03; ▲: 0.045; ▼: 0.06), while the broken line is for films of a mixture of *T*- and *T'*-phase (○: 0.09).

Fig. 5. Variation of resistivity at 300 K [$\rho$(300 K)] for *T*-$La_{2-x}Ce_xCuO_4$ (●), *T*-$La_{2-x}Sr_xCuO_4$ (○), and *T'*-$La_{2-x}Ce_xCuO_4$ (□).

Table I. The $a$-axis lattice constants ($a_s$) for the substrates used in this work.

| Substrate | Abbreviation | $a_s$ (Å) |
|---|---|---|
| $LaSrGaO_4$ | LSGO | 3.843 |
| $LaSrAlO_4$ | LSAO | 3.755 |
| $PrSrAlO_4$ | PSAO | 3.727 |
| $NdSrAlO_4$ | NSAO | 3.712 |
| $NdCaAlO_4$ | NCAO | 3.688 |

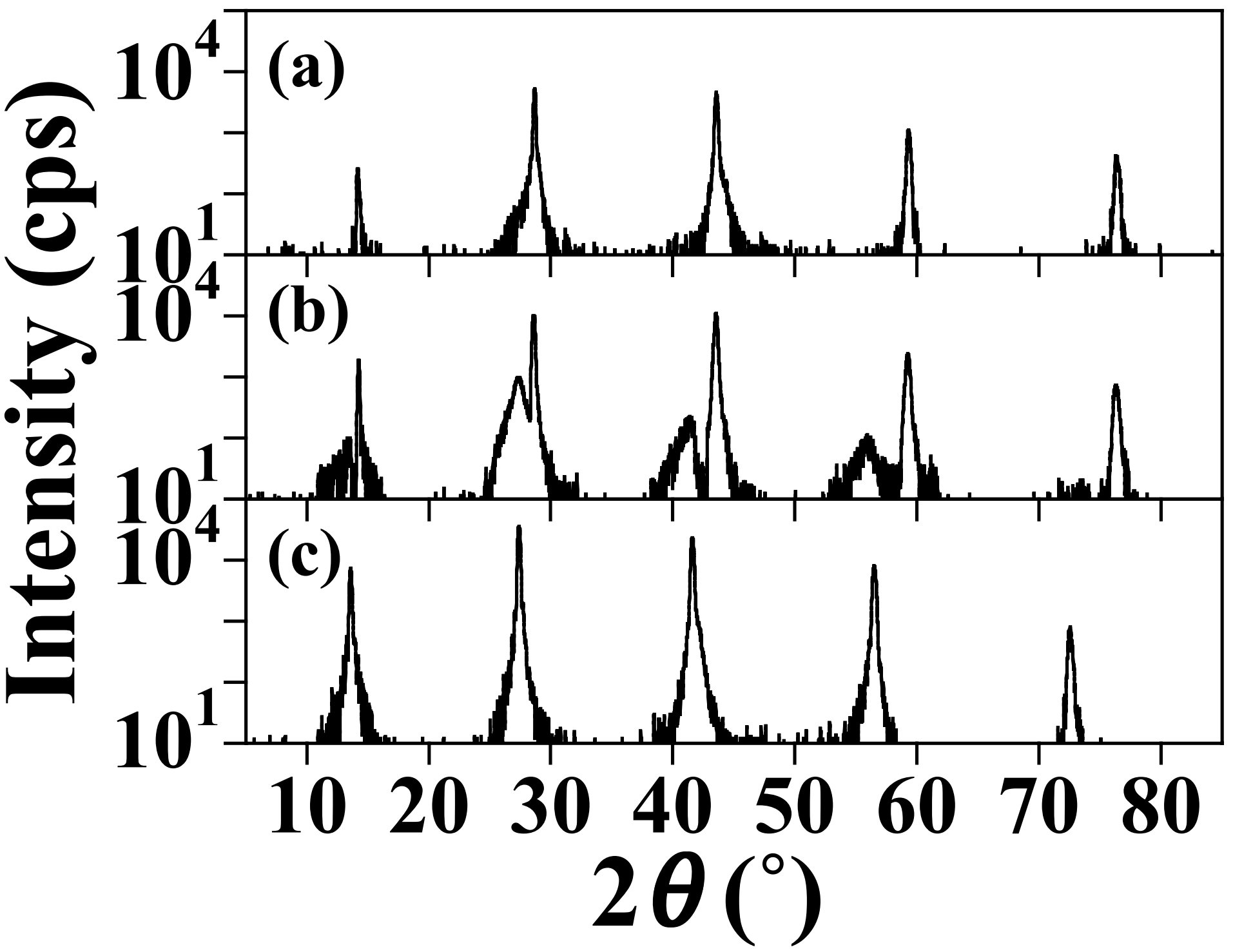

**Figure 1, Tsukada *et al.***

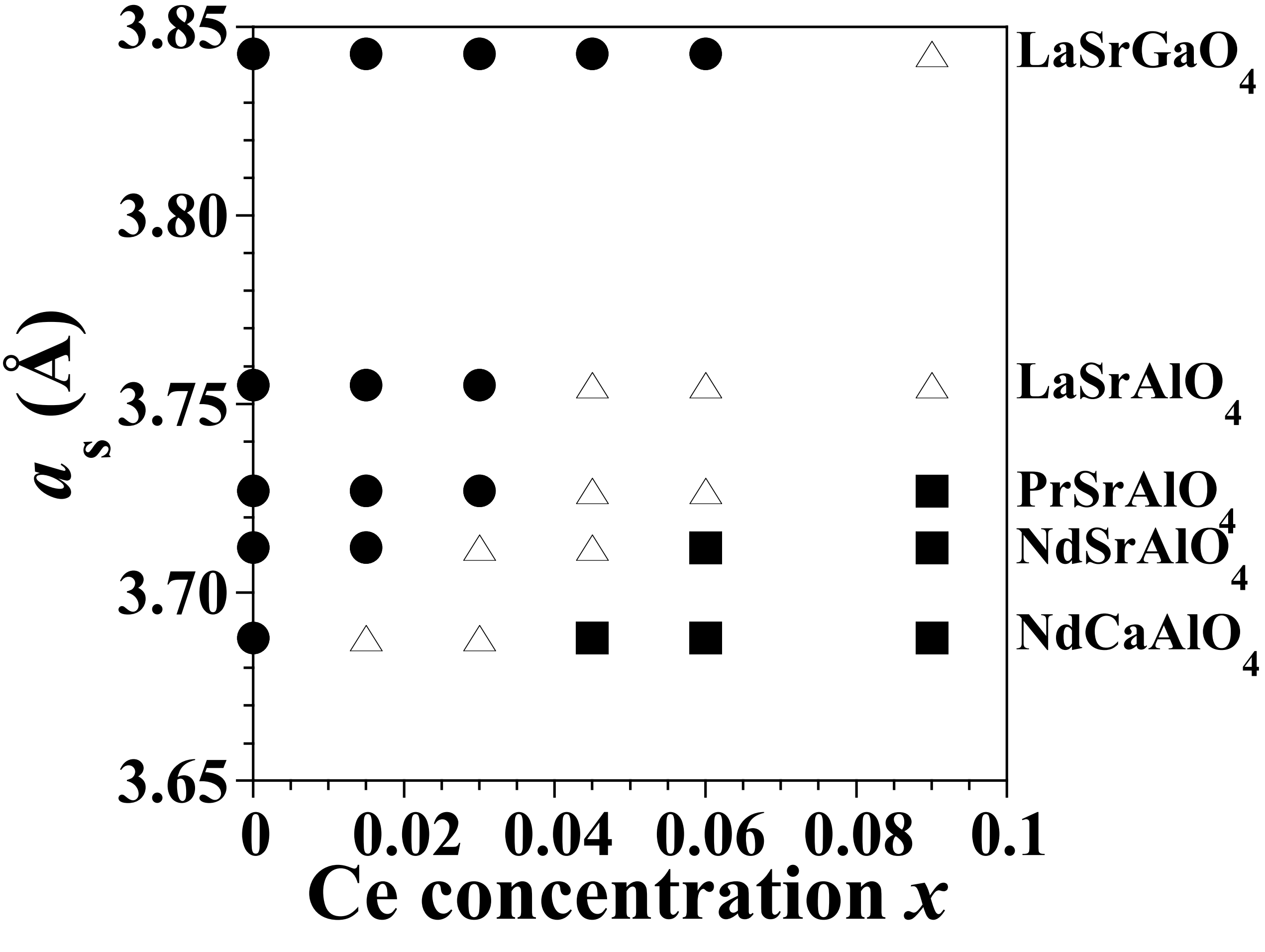

**Figure 2, Tsukada *et al.***

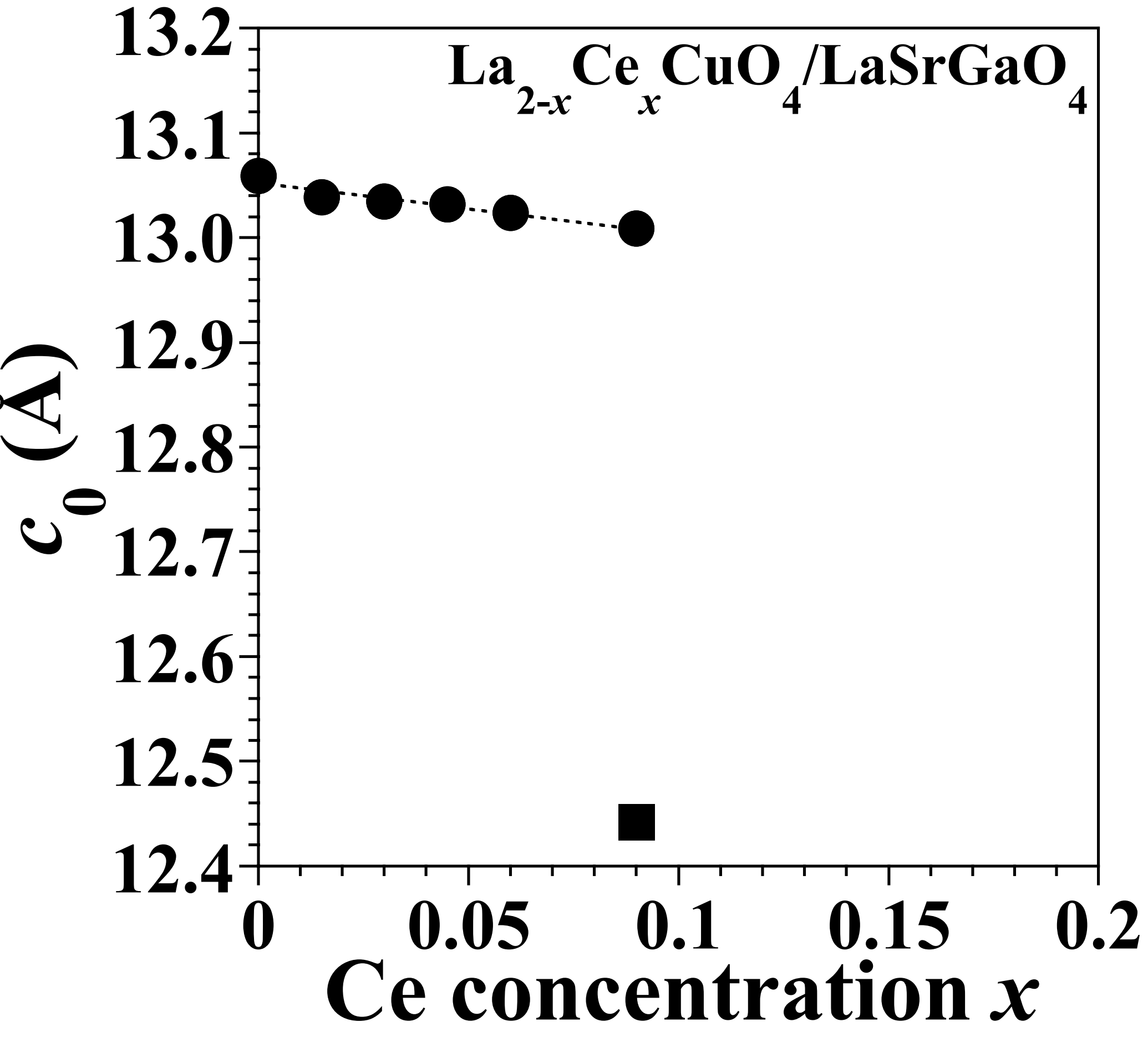

**Figure 3, Tsukada *et al.***

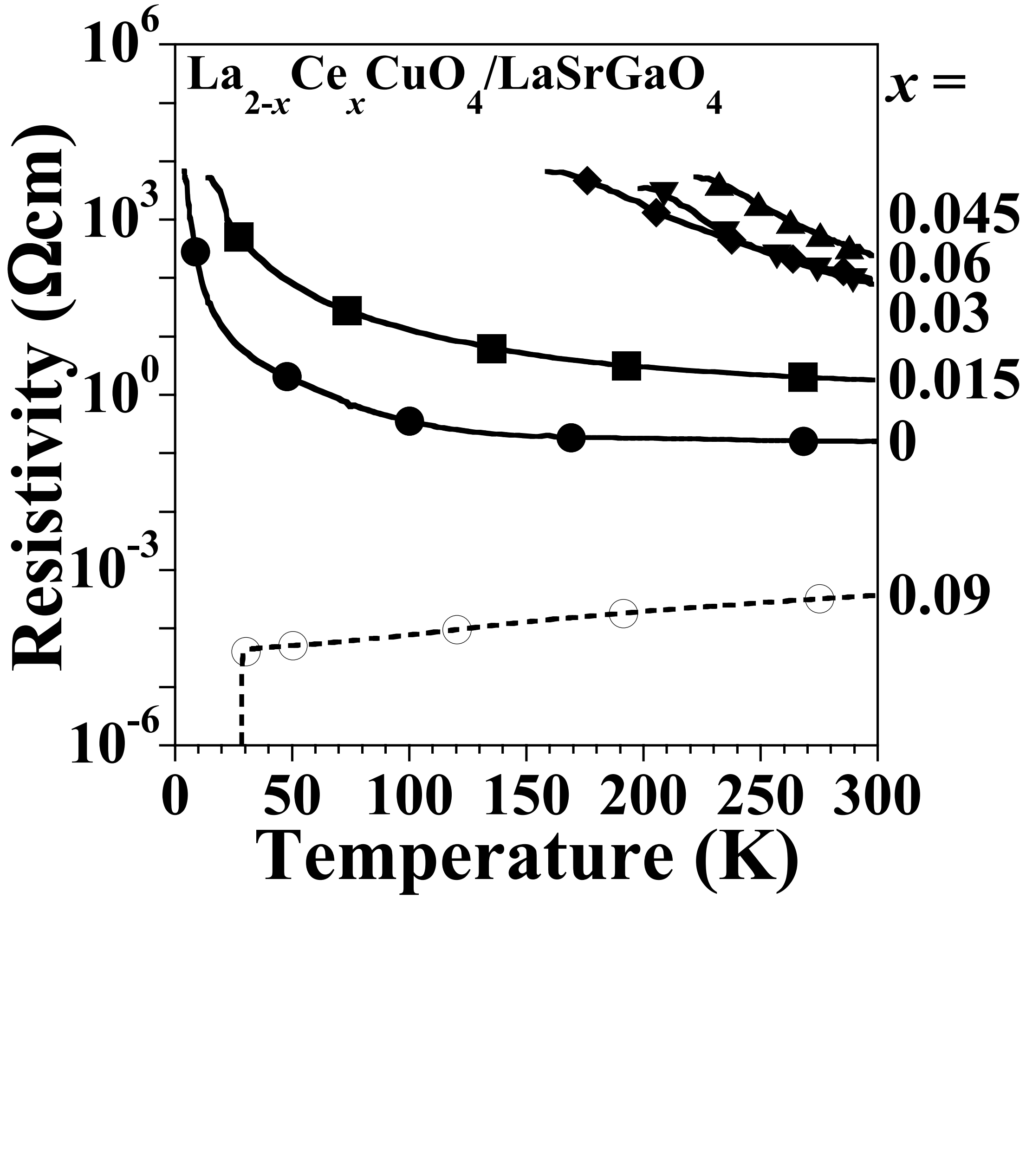

**Figure 4, Tsukada *et al.***

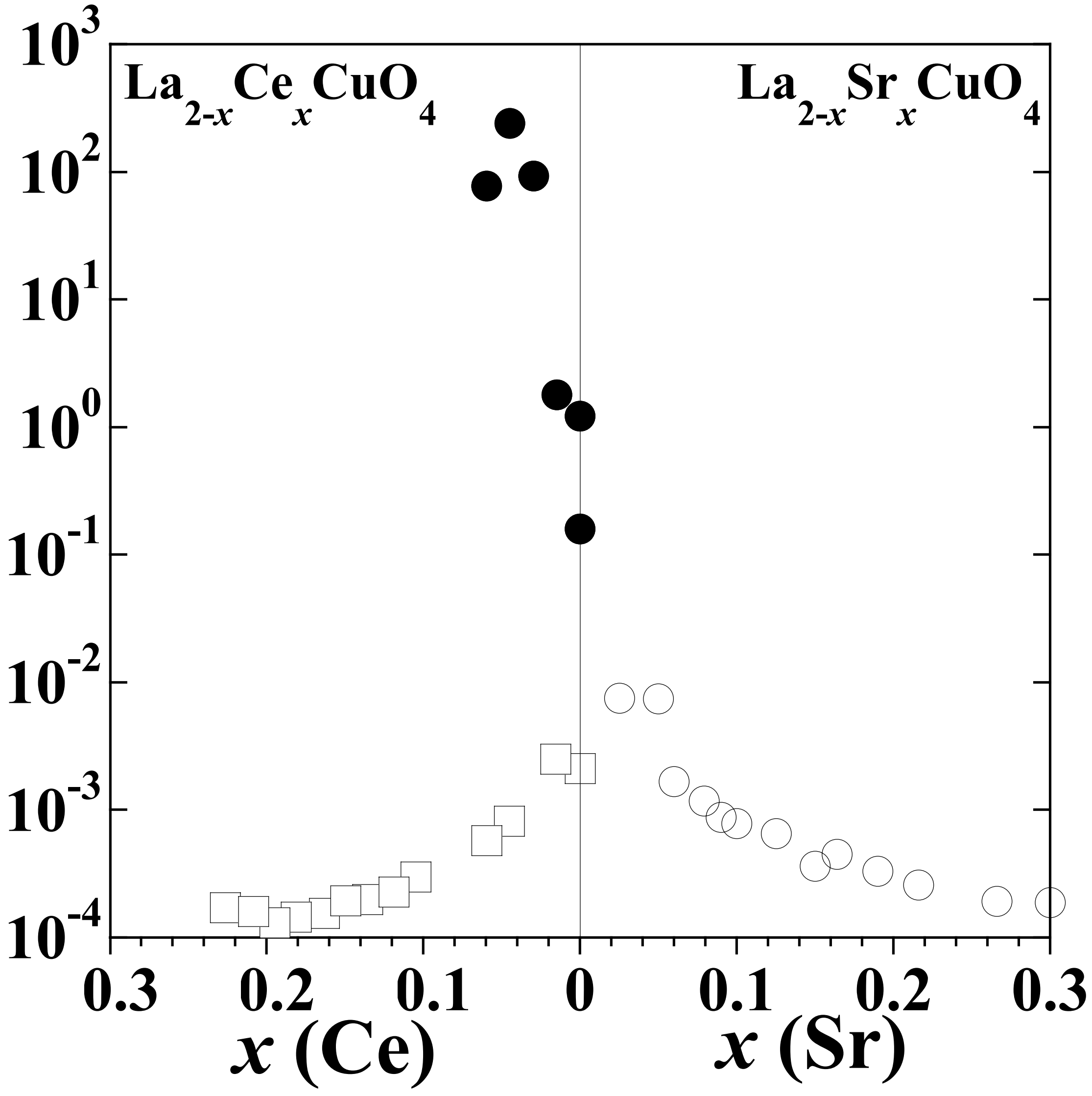

**Figure 5, Tsukada *et al.***